\documentclass[aps,prd,twocolumn,showpacs,amsmath,amssymb,nofootinbib]{revtex4}
\usepackage{graphicx}
\usepackage{dcolumn}
\usepackage{bm}
\begin{document}
\title{Generalized Tensor Analysis Method Applied to\\Non-time-orthogonal Coordinate Frames}
\author{Robert D. Klauber}
\affiliation{  1100 University Manor Dr., 38B\\Fairfield, Iowa 52556}
\email{rklauber@.netscape.net}
\date{August 22, 2003}
            
\begin{abstract}

A generalized covariant method of analysis applicable to coordinate frames for which 
time is not orthogonal to space, such as spacetime around a star possessing 
angular momentum or on a rotating disk, is presented. Important aspects of 
such an analysis are shown to include i) use of the physically relevant 
contravariant or covariant component form for a given vector/tensor, ii) 
conversion of physical (measured) components to generalized coordinate 
components prior to tensor analysis, iii) use of generalized covariant 
constitutive equations during tensor analysis, and iv) conversion of 
coordinate components back to physical components after tensor analysis. The 
method is applied to electrodynamics in a rotating frame, and shown to 
predict the measured results of the Wilson and Wilson and 
Roentgen/Eichenwald experiments.

\end{abstract}
\maketitle

\section{INTRODUCTION}
The literature on electrodynamic analysis in rotating frames, though fairly 
extensive, rarely addresses the issue of relating generalized coordinate 
values for electric and magnetic field strengths to those physical values 
measured in experiment. When it is addressed, ambiguity seems to arise.

For example, Crater\cite{Crater:1994}, advocates ``apparent freedom of 
choice for which components of the electromagnetic second-rank $F$ tensor, 
(contravariant, covariant, or mixed) one uses...'' and that on the rotating 
frame, ``.. one would have two devices for measuring voltages, capacitances, 
inductances, etc., one for those physicists who prefer the covariant 
convention and one for those who prefer the contravariant convention.'' This 
begs the question, which Crater does not bring up: What if one simply took 
the usual volt meter used in the lab onto the rotating frame. Which of these 
various coordinate types, if either, would it read?

As another example, Ridgely\cite{Ridgely:1999} notes that contravariant 
electromagnetic field tensors have a vector basis and covariant tensors have 
a one-form basis, but does not suggest which of them corresponds to values 
read on instruments located in a rotating frame. Nor does he consider the 
closely related issue of determining physical (measured) components from 
coordinate (mathematical) components (be they contravariant or covariant). 
The present article answers these questions and presents a general analysis 
method suitable for, but not limited to, electrodynamics in rotating frames. 

In Section \ref{sec:mylabel1} a generalized tensor analysis method 
for mechanics and/or electrodynamics is developed, that while applicable to 
any frame, finds specific utility for those coordinate frames in which time 
is not orthogonal to space. Such coordinate frames have off diagonal 
space-time terms in the metric, and are designated ``non-time-orthogonal'' 
(NTO). They include rotating frames, as well as spacetime around objects 
having significant angular momentum.

Section \ref{sec:rotating} provides a brief mathematical review of 
rotating frames. That, along with the method of Section 
\ref{sec:mylabel1}, is then applied in Section 
\ref{sec:wilson} to the Wilson and Wilson experiment, and in 
Section \ref{sec:mylabel2} to the Roentgen/Eichenwald experiment. It 
is shown that the NTO nature of the rotating frame introduces terms into the 
analysis that would not be present in a time orthogonal (TO) frame analysis, 
and that agreement between theory and experiment is directly attributable to 
those additional terms.

\section{NTO TENSOR ANALYSIS: RULES OF THE GAME}
\label{sec:mylabel1}
\subsection{Contravariant or covariant?}
\label{subsec:contravariant}
Generalized position coordinates $x^{\mu }$ are contravariant in nature. That 
is, the generalized displacement four-vector between two infinitesimally 
separated 4D points (events) is, to be precise, \textit{dx}$^{\mu }$, not \textit{dx}$_{\mu }$. In 
the (special) case where basis vectors \textbf{e}$_{\mu }$ are orthogonal 
(i.e., the coordinate grid lines are all orthogonal to one another), this 
distinction is not critical. (See Appendix A.) In the more general case 
(e.g., the NTO case), however, where all basis vectors \textbf{e}$_{\mu }$ 
are not orthogonal to one another, the distinction between \textit{dx}$^{\mu }$ and 
\textit{dx}$_{\mu }$ becomes important, and one must keep in mind that they do not 
represent the same entity in the physical world. In an NTO frame, 
\textit{dx}$^{{\rm 0}}$ corresponds to displacement along the time axis, whereas 
\textit{dx}$_{0}$ corresponds to displacement in both time and space. Hence, in the 
most general and accurate sense, \textit{dx}$^{\mu }$ (and not \textit{dx}$_{\mu })$ is the 
correct representation of coordinate displacement.

The need to distinguish between the physical significance of contravariant 
and covariant components carries over to other four vectors. For example, 
the four velocity
\begin{equation}
\label{eq1}
u^\mu =\frac{dx^\mu }{d\tau }
\end{equation}
must be a contravariant vector since \textit{dx}$^{\mu }$ is contravariant. As before, 
the covariant vector components $u_\mu =g_{\mu \upsilon } u^\upsilon $ may 
be considered an equivalent form of the four velocity in an \textit{orthogonal} coordinate 
system (such as Minkowski coordinates in a Lorentz frame), but not in the 
more general NTO case.

Four momentum is defined, where $L$ is the Lagrangian, as
\begin{equation}
\label{eq2}
p_\mu =\frac{\partial L}{\partial u^\mu }
\end{equation}
and hence, it must be covariant since $u^{\mu }$ is contravariant. For a free 
particle $L=\textstyle{1 \over 2}mu^\mu u_\mu $ and four-momentum is $mu_\mu 
$, not $mu^\mu $. In an NTO frame these quantities are distinctly different, 
and in fact, as the author has shown\cite{Klauber:1998}, the correct value 
for energy in an NTO rotating frame is found from $p_{{\rm 0}}$ in the 
rotating frame, not $p^{0}$.

The four current density $J^{\mu }$ obeys the physical law of conservation of 
charge
\begin{equation}
\label{eq3}
J_{\,\,\,\,\,;\mu }^\mu =0,
\end{equation}
where we have generalized to non-Minkowski coordinate systems via the 
covariant derivative symbolized by the semi-colon. Since the covariant 
derivative is with respect to the (contravariant) four vector displacement, 
the only possibility whereby (\ref{eq3}) can equal a 4D scalar invariant (zero) is 
if the physical world four current density is represented in contravariant 
form. Certainly $J_\mu ^{\,\,\,\,;\mu } =0$ holds, but in the most general 
case, the contravariant derivative $\partial ^0$ (raised ;$\mu $ is with 
respect to \textit{dx}$_{\mu })$ is not a derivative with respect solely to time, but 
rather with respect to an amalgam of both space and time. Hence 
$J_0^{\,\,\,;0} $ would not equal the time rate of change of charge density 
we would measure with physical instruments, and $J_{0}$ would not correspond 
to charge density, even in a generalized sense.

In general, we can use physical laws, which hold covariantly in a 
generalized sense, to determine what form (covariant or contravariant) any 
given vector or tensor must be represented by in order for it to correspond 
to a physical world entity. This is, in fact, what we did with four current 
and the charge conservation law above. As a second example, four-force is 
the derivative of four momentum with respect to proper time and so must be 
covariant\footnote{ Care must be taken, however, if the four force is to be 
related to motion. That is, $p^{\mu }$ = \textit{mu}$^{\mu }$ must be found from 
$f^{\mu }$, not $f_{\mu }$ , and $u^{\mu }$ integrated with respect to proper 
time to find coordinate displacement $\Delta x^{\mu }$.}.

In electrodynamics, one can use Maxwell's equations and the Lorentz force 
law formulated in 4D to determine the appropriate form of field tensors. In 
a Minkowski coordinate system the 3D electric and magnetic fields 
\textbf{E}, \textbf{B}, \textbf{D}, and \textbf{H} form the 4D 
tensors\cite{Jackson:1975}$^{,}$\cite{Misner:1973}
\begin{eqnarray}
\label{eq4}
F^{\mu \upsilon }=\left( {{\begin{array}{*{20}c}
 0 \hfill & {E^1} \hfill & {E^2} \hfill & {E^3} \hfill \\
 {-E^1} \hfill & 0 \hfill & {B^3} \hfill & {-B^2} \hfill \\
 {-E^2} \hfill & {-B^3} \hfill & 0 \hfill & {B^1} \hfill \\
 {-E^3} \hfill & {B^2} \hfill & {-B^1} \hfill & 0 \hfill \\
\end{array} }} \right)\; \nonumber \\*
H^{\mu \upsilon }=\left( 
{{\begin{array}{*{20}c}
 0 \hfill & {D^1} \hfill & {D^2} \hfill & {D^3} \hfill \\
 {-D^1} \hfill & 0 \hfill & {H^3} \hfill & {-H^2} \hfill \\
 {-D^2} \hfill & {-H^3} \hfill & 0 \hfill & {H^1} \hfill \\
 {-D^3} \hfill & {H^2} \hfill & {-H^1} \hfill & 0 \hfill \\
\end{array} }} \right).
\end{eqnarray}
The generalized Maxwell source equations in 4D for Gaussian units are
\begin{equation}
\label{eq5}
H_{\,\,\,\,\,\,\,;\nu }^{\mu \nu } =\frac{4\pi }{c}J^\mu ,
\end{equation}
where $H^{\mu \nu }=F^{\mu \nu }$ in vacuum. Since the derivative is 
covariant (lowered index) in form and the current density is contravariant, 
the $H$ tensor (and hence also the $F$ tensor) dependent on (arising from) charge 
and current sources must be contravariant. 

Note, however, that finding the four-force (covariant) vector acting on a 
charge or three-current density is determined via
\begin{equation}
\label{eq6}
f_\mu =\frac{1}{c}F_{\mu \nu } J^\nu .
\end{equation}
Hence, since four-force is covariant and four-current is contravariant, the 
$F$ tensor must, in this case, be covariant. We conclude that $F^{\mu \nu }$ 
represents the physical fields \textit{found from} charges [see (\ref{eq5})] and currents in a vacuum, 
but $F_{\mu \nu }$ represents the physical fields as they \textit{act on} charges and 
currents [see (\ref{eq6})] to produce forces. Again, this distinction is not 
critical in TO frames, though it is quite relevant in NTO frames when one 
wishes to know which electric and magnetic field components (contravariant 
or covariant) will be monitored by physical instruments. Since the covariant 
form of $F$ represents the physical forces acting on charges and currents 
(which form the basis of field measuring devices such as voltmeters and 
Gauss meters), those covariant components $F_{\mu \nu }$ correlate with what 
one would measure in an experiment. [However, one must ascertain that the 
principle of operation of the measuring device is force based (covariant 
$f_{\mu})$ rather than motion based (contravariant $m\dot {u}^\mu $ on the 
LHS of (\ref{eq6}).)]

\subsection{Physical vs. Coordinate Components}
\label{subsec:physical}
Getting the correct contravariant or covariant components is not quite 
enough, however, in order to compare theoretical results with measured 
quantities. A generalized vector component (e.g., four velocity component 
$u^{1})$ is a mathematical entity whose value generally does not equal the 
value one would measure with physical instruments (e.g., the velocity 
measured using standard rods and clocks in the $x^{1}$ axis direction.)

If a given basis vector does not have unit length, the magnitude of the 
associated component will not equal the physical quantity measured. For 
example, a vector with a single non-zero component value of 1 in a 
coordinate system where the corresponding basis vector for that component 
has length 3 does not have an absolute (physical) length equal to 1, but to 
three. In general, the \textit{physical} component (measured) value is equal to the 
generalized coordinate component value only in the special case where the 
basis vector has unit length (is normalized.) One example of this is Lorentz 
frames with Minkowski coordinates, which have unit basis vectors and 
vector/tensor components equal to the physically measured values.

To determine physical component values, we need to calculate, given the 
generalized coordinate component $v^{\alpha }$, what the component value 
would be for a unit length basis vector. In Appendix B, for reference, we 
derive this well-known relation\cite{Ref:1988}, i.e., 
\begin{equation}
\label{eq7}
\begin{array}{l}
 v^{\hat {i}}=\sqrt {g_{\underline{i}\underline{i}} } v^i\quad \quad v^{\hat 
{0}}=\sqrt {-g_{00} } v^0 \\ 
 v_{\hat {i}} =\sqrt {g^{\underline{i}\underline{i}}} v_i \quad \quad 
v_{\hat {0}} =\sqrt {-g^{00}} v_0 \\ 
 \end{array},
\end{equation}
where underlining implies no summation, carets over indices indicate 
physical components, spatial coordinates are designated by Roman script, and 
the minus signs are needed as $g^{00}$ and $g_{00}$ are negative.

This result is readily extended to second order tensors, i.e.,
\begin{equation}
\label{eq8}
T^{\hat {\mu }\hat {\upsilon }}=\sqrt {g_{\underline{\mu }\underline{\mu }} 
} \sqrt {g_{\underline{\upsilon }\underline{\upsilon }} } T^{\mu \upsilon 
}\quad \quad T_{\hat {\mu }\hat {\upsilon }} =\sqrt {g^{\underline{\mu 
}\underline{\mu }}} \sqrt {g^{\underline{\upsilon }\underline{\upsilon }}} 
T_{\mu \upsilon } 
\end{equation}
where, to be precise, multiplication of metric components by (-1) is 
required whenever they are negative.

We note that physical components are a special case of anholonomic 
components\footnote{ Anholonomic, or non-coordinate, components are those 
components associated with non-coordinate basis vectors. For the special 
case where these non-coordinate basis vectors have unit length, anholonomic 
components equal physical components.}\cite{Ref:1971}. We caution 
that physical components do not transform according to the vector/tensor 
transformation laws and are not components of vectors/tensors\cite{Ref:1}. 
Hence we may calculate physical components to determine what we would 
measure via experiment, but we must use coordinate components in carrying 
out tensor analysis.

We note further that physical components $v^{\hat {\mu }}$ and $v_{\hat {\mu 
}} $ are identical in orthogonal axes systems, but are generally different 
in non-orthogonal axis systems. This underscores the theme of Section 
\ref{subsec:contravariant} and the need to find physical components for 
the physically relevant (contravariant or covariant) tensor form in order to 
compare theory with experiment in an NTO frame.

\subsection{Covariant Constitutive Equations}
\label{subsec:covariant}
The 3D plus time constitutive relations
\begin{equation}
\label{eq9}
{\rm {\bf D}}=\varepsilon {\rm {\bf E}}\quad \quad \quad {\rm {\bf B}}=\mu 
{\rm {\bf H}}
\end{equation}
were expressed by Minkowski in the 4D covariant form\cite{Landau:1984}
\begin{equation}
\label{eq10}
H^{\mu \upsilon }u_\upsilon =\varepsilon F^{\mu \upsilon }u_\upsilon 
\end{equation}
\begin{equation}
\label{eq11}
\varepsilon _{\sigma \lambda \mu \nu } F^{\lambda \mu }u^\upsilon =\mu 
\varepsilon _{\sigma \lambda \mu \nu } H^{\lambda \mu }u^\upsilon .
\end{equation}
We will use (\ref{eq10}) and (\ref{eq11}) to relate $F^{\mu \nu }$ to $H^{\mu \nu }$ in NTO 
frames.

\subsection{General Method of Analysis}
\label{subsec:general}
Analysis of the most general type of problem therefore comprises the 
following steps.

1. Determine whether known (measured) component values are contravariant or 
covariant in physical character. [See (\ref{eq1}) through (\ref{eq6}).]

2. Convert the appropriate contravariant or covariant physical components to 
the corresponding coordinate components via (\ref{eq7}) and/or (\ref{eq8}).

3. Apply tensor analysis as relevant to the particular problem, using 
generalized covariant constitutive equations.

4. Convert the coordinate component (contravariant or covariant as 
physically appropriate) answer to physical components to determine what 
would be measured in the physical world.

\section{ROTATING FRAME TRANSFORMATION}
\label{sec:rotating}
For the rotating frame analysis we employ cylindrical coordinates with 
(\textit{cT,R,$\Phi $,Z) }for the lab and (\textit{ct,r,}$\phi $,$z)$ for the rotating frame. The transformation 
below, between the lab and a rotating frame having angular velocity $\omega 
$ in the Z direction, is well known, widely used, and argued elsewhere by 
the 
author\cite{Klauber:1}$^{,}$\cite{Klauber:1999}$^{,}$\cite{Klauber:2003}$^{,}$\cite{Klauber:1998} 
to be appropriate for analyses such as that herein.
\begin{equation}
\label{eq12}
\begin{array}{l}
 cT=ct \\ 
 R=r \\ 
 \Phi =\phi +\omega t \\ 
 Z=z\,. \\ 
 \end{array}
\end{equation}
In matrix form this may be expressed as
\begin{equation}
\label{eq13}
\Lambda _{\,\,\,B}^\alpha =\left[ {{\begin{array}{*{20}c}
 1 \hfill & 0 \hfill & 0 \hfill & 0 \hfill \\
 0 \hfill & 1 \hfill & 0 \hfill & 0 \hfill \\
 {-\textstyle{\omega \over c}} \hfill & 0 \hfill & 1 \hfill & 0 \hfill \\
 0 \hfill & 0 \hfill & 0 \hfill & 1 \hfill \\
\end{array} }} \right]\quad \quad \Lambda _{\,\,\,\beta 
}^A =\left[ {{\begin{array}{*{20}c}
 1 \hfill & 0 \hfill & 0 \hfill & 0 \hfill \\
 0 \hfill & 1 \hfill & 0 \hfill & 0 \hfill \\
 {\textstyle{\omega \over c}} \hfill & 0 \hfill & 1 \hfill & 0 \hfill \\
 0 \hfill & 0 \hfill & 0 \hfill & 1 \hfill \\
\end{array} }} \right],
\end{equation}
where $A$ and $B$ here are upper case Greek, $\Lambda _{\,\,\,B}^\alpha $ 
transforms a contravariant lab vector [e.g., \textit{dX}$^{B}$ = (\textit{cdT,dR,d}$\Phi $\textit{,dZ})$^{T}$] to 
the corresponding contravariant rotating frame vector [\textit{dx}$^{\alpha }$ = 
(\textit{cdt,dr,d}$\phi $\textit{,dz})$^{T}$], and $\Lambda _{\,\,\,\beta }^A $ transforms the latter 
back from the rotating frame to the lab.

The following relations, which we will use in Sections IV and V, 
are derived in Klauber\cite{Ref:2} from (\ref{eq12}). The rotating frame coordinate 
metric $g_{\alpha \beta }$ and its inverse $g^{\alpha \beta }$ are
\begin{eqnarray}
\label{eq14}
g_{\alpha \beta } =\left[ {{\begin{array}{*{20}c}
 {-(1-\textstyle{{r^2\omega ^2} \over {c^2}})} \hfill & 0 \hfill & 
{\textstyle{{r^2\omega } \over c}} \hfill & 0 \hfill \\
 0 \hfill & 1 \hfill & 0 \hfill & 0 \hfill \\
 {\textstyle{{r^2\omega } \over c}} \hfill & 0 \hfill & {r^2} \hfill & 0 
\hfill \\
 0 \hfill & 0 \hfill & 0 \hfill & 1 \hfill \\
\end{array} }} \right]  \nonumber  \\*
 g^{\alpha \beta 
}=\left[ {{\begin{array}{*{20}c}
 {-1} \hfill & 0 \hfill & {\textstyle{\omega \over c}} \hfill & 0 \hfill \\
 0 \hfill & 1 \hfill & 0 \hfill & 0 \hfill \\
 {\textstyle{\omega \over c}} \hfill & 0 \hfill & 
{\textstyle{{(1\;\;-\;\;\textstyle{{r^2\omega ^2} \over {c^2}})} \over 
{r^2}}} \hfill & 0 \hfill \\
 0 \hfill & 0 \hfill & 0 \hfill & 1 \hfill \\
\end{array} }} \right].
\end{eqnarray}
The lab metric $G_{AB}$ and its inverse$ G^{AB}$ are
\begin{equation}
\label{eq15}
G_{AB} =\left[ {{\begin{array}{*{20}c}
 {-1} \hfill & 0 \hfill & 0 \hfill & 0 \hfill \\
 0 \hfill & 1 \hfill & 0 \hfill & 0 \hfill \\
 0 \hfill & 0 \hfill & {R^2} \hfill & 0 \hfill \\
 0 \hfill & 0 \hfill & 0 \hfill & 1 \hfill \\
\end{array} }} \right]\quad G^{AB}==\left[ 
{{\begin{array}{*{20}c}
 {-1} \hfill & 0 \hfill & 0 \hfill & 0 \hfill \\
 0 \hfill & 1 \hfill & 0 \hfill & 0 \hfill \\
 0 \hfill & 0 \hfill & {\textstyle{1 \over {R^2}}} \hfill & 0 \hfill \\
 0 \hfill & 0 \hfill & 0 \hfill & 1 \hfill \\
\end{array} }} \right].
\end{equation}
\section{WILSON AND WILSON RESULT}
\label{sec:wilson}
The Wilson and Wilson\cite{Wilson:1913} experiment comprised a rotating 
cylinder of magnetic permeability $\mu $ and dielectric constant 
$\varepsilon $ and an axially directed uniform magnetic field $B_{0}$. They 
measured a radially directed electric field (consonant with a magnetic field 
internal to the cylinder) in the lab of
\begin{equation}
\label{eq16}
E_{\scriptsize \begin{array}{l}
 R \\ 
 lab,int \\ 
 measured \\ 
 \end{array}} =\frac{vB_0 }{c}\left( {\frac{1}{\varepsilon }-\mu } \right),
\end{equation}
where $v=\omega r$.

Recently, Pellegrini and Swift\cite{Pellegrini:1995} (PS) seemed to show 
that, based on the global coordinate transformation [(\ref{eq12}) in Cartesian, 
rather than cylindrical, form] to a rotating frame, the theoretical 
prediction for the Wilsons experiment disagreed with the actual test 
results. Subsequent testing by Hertzberg et al\cite{Hertzberg:2001} 
confirmed the validity of the Wilson and Wilson result. As shown by PS, and 
in greater detail by Weber\cite{Weber:1997}, the correct answer could be 
found by using local Lorentz frames as surrogates for the global rotating 
frame.

PS noted their global transformation to the rotating frame resulted in a 
metric with off diagonal space-time components (i.e., an NTO metric.) PS 
suggested that the resolution to the issue might lie in the form taken by 
the constitutive equations in the NTO rotating frame. Subsequently, 
covariant expressions for the constitutive equations were found and used by 
Burrows\cite{Burrows:1997} and Ridgely\cite{Ridgely:1998} to yield the 
correct prediction for the Wilsons experiment.

However, Burrows and Ridgely carried out much of their analyses with 
Maxwell's equations expressed in 3D form, and did not use a fully 4D 
generalized covariant tensor method throughout. Further, Ridgely did not 
employ the widely accepted transformation to the rotating frame used by PS, 
leaving unanswered the question of the ultimate validity of that 
transformation. Still further, PS showed that with certain assumptions they 
could derive the Wilsons result, yet with those same assumptions, they 
arrived at an incorrect prediction for a different experiment performed by 
both Roentgen\cite{Roentgen:1} and Eichenwald\cite{Eichenwald:1904} (RE). 
Neither Burrows nor Ridgely\footnote{ In fact, in ref. 
\cite{Ridgely:1998}, p. 119 just prior to equation (45a), Ridgely states 
that for an axially directed laboratory electric field, the magnetic 
induction field \textbf{B}$^{/}$ = \textbf{0} in the rotating frame. This 
appears to contradict the experimental result of Roentgen and Eichenwald, 
refs. \cite{Roentgen:1} and \cite{Eichenwald:1904}, and stated in ref. 
\cite{Pellegrini:1995}, p. 700 equation (\ref{eq27}) and following paragraph 
therein.} addressed the RE experiment directly, leaving open the question as 
to what their analyses would predict for that test.

In the following sections we employ the steps of section 
\ref{subsec:general} to derive (\ref{eq16}). We also use those steps in 
Appendix C where the issue of modeling of the magnetization by surface 
currents as described in PS is resolved.

\subsection{Overview of Analysis Procedure for Wilsons Experiment}
\label{subsec:overview}
Steps 1 and 2: Convert the physical component $B_{0}$ measured in the lab in 
the air to the associated covariant component of the 4D $F$ tensor. Obtain the 
corresponding contravariant form of the $F$ tensor.

Step 3: Transform the $F$ tensor to the rotating frame. Take the $F$ tensor as 
equal to the $H$ tensor in air. Apply boundary conditions obtained from the 4D 
Maxwell equations at the boundary of the cylinder to find the $H$ tensor inside 
the rotating cylinder. Apply the constitutive equations (\ref{eq10}) and (\ref{eq11}) to the 
$H$ tensor to obtain the $F$ tensor inside the cylinder. Transform the $F$ tensor back 
to the lab. 

Step 4: Convert the appropriate component of $F$ to obtain the physical 
component in the radial direction of the lab electric field inside the 
cylinder as in (\ref{eq16}).

Schematically, the individual mathematical steps in the analysis are 
outlined in (\ref{eq17}) to (\ref{eq19}). ``Ext'' means external to the rotating permeable 
dielectric cylinder, and ``Int'' means internal to the cylinder.
\begin{eqnarray}
\label{eq17}
\mbox{phys comps}\,\,\,F_{\scriptsize \begin{array}{l}
 \hat {A}\hat {B} \\ 
 lab,ext \\ 
 \end{array}} \;\to F_{\scriptsize \begin{array}{l}
 AB \\ 
 lab,ext \\ 
 \end{array}} \to F_{lab,ext}^{AB} \nonumber  \\*
\quad \quad \quad \to F_{rot,ext}^{\mu \nu } 
=H_{rot,ext}^{\mu \nu } 
\end{eqnarray}
Then use the source Maxwell equations (\ref{eq5}) with zero source term and the 4D 
Gauss divergence law to apply boundary conditions at the rotating material 
boundary between the external and internal \textbf{D} and \textbf{H} fields. 
That is,
\begin{eqnarray}
\label{eq18}
H_{\,\,\,\,\,\,\,;\mu }^{\mu \nu } =0\;\quad \to \quad n_\mu  H_{rot,ext}^{\mu \nu } =n_\mu H_{rot,int}^{\mu \nu } & &    \nonumber \\*
 \to \quad 
H_{rot,ext}^{\mu \nu } =H_{rot,int}^{\mu \nu } &
\end{eqnarray}
Next apply the Minkowski covariant constitutive relations of (\ref{eq10}) and (\ref{eq11}) 
to get $F_{rot,int}^{\mu \upsilon } $, and transform back to the lab, i.e.,
\begin{eqnarray}
\label{eq19}
H_{rot,int}^{\mu \nu } \quad \to \quad F_{rot,int}^{\mu \nu } \quad \to 
\quad F_{lab,int}^{AB} \quad \to \quad F_{\scriptsize \begin{array}{l}
 AB \\ 
 lab,int \\ 
 \end{array}}  \nonumber \\*
\quad  \quad  \quad \to \quad \mbox{phys comps}\,\,\,F_{\scriptsize \begin{array}{l}
 \hat {A}\hat {B} \\ 
 lab,int \\ 
 \end{array}} .
\end{eqnarray}
\subsection{Application of Above Procedure}
\label{subsec:application}
The lab physical component \textbf{E},\textbf{B} field 4D ``tensor'' for an 
axially directed magnetic field $B_{0}$ is
\begin{equation}
\label{eq20}
F_{\scriptsize \begin{array}{l}
 \hat {A}\hat {B} \\ 
 lab,ext \\ 
 \end{array}} =\left[ {{\begin{array}{*{20}c}
 0 \hfill & 0 \hfill & 0 \hfill & 0 \hfill \\
 0 \hfill & 0 \hfill & {B_0 } \hfill & 0 \hfill \\
 0 \hfill & {-B_0 } \hfill & 0 \hfill & 0 \hfill \\
 0 \hfill & 0 \hfill & 0 \hfill & 0 \hfill \\
\end{array} }} \right]
\end{equation}
where
\begin{equation}
\label{eq21}
F_{\scriptsize \begin{array}{l}
 \hat {R}\hat {\Phi } \\ 
 lab,ext \\ 
 \end{array}} =\sqrt {G^{RR}} \sqrt {G^{\Phi \Phi }} F_{\scriptsize \begin{array}{l}
 R\Phi \\ 
 lab,ext \\ 
 \end{array}} =\frac{1}{R}F_{\scriptsize \begin{array}{l}
 R\Phi \\ 
 lab,ext \\ 
 \end{array}} ,
\end{equation}
so the coordinate component \textbf{E},\textbf{B} field 4D covariant tensor 
in the lab is
\begin{equation}
\label{eq22}
F_{\scriptsize \begin{array}{l}
 AB \\ 
 lab,ext \\ 
 \end{array}} =\left[ {{\begin{array}{*{20}c}
 0 \hfill & 0 \hfill & 0 \hfill & 0 \hfill \\
 0 \hfill & 0 \hfill & {RB_0 } \hfill & 0 \hfill \\
 0 \hfill & {-RB_0 } \hfill & 0 \hfill & 0 \hfill \\
 0 \hfill & 0 \hfill & 0 \hfill & 0 \hfill \\
\end{array} }} \right].
\end{equation}
The contravariant form, needed to transform to the rotating frame via (\ref{eq13}), 
is
\begin{equation}
\label{eq23}
F_{lab,ext}^{AB} =G^{A\Gamma }G^{B\Omega }F_{\scriptsize \begin{array}{l}
 \Gamma \Omega \\ 
 lab,ext \\ 
 \end{array}} =\left[ {{\begin{array}{*{20}c}
 0 \hfill & 0 \hfill & 0 \hfill & 0 \hfill \\
 0 \hfill & 0 \hfill & {\frac{B_0 }{R}} \hfill & 0 \hfill \\
 0 \hfill & {-\frac{B_0 }{R}} \hfill & 0 \hfill & 0 \hfill \\
 0 \hfill & 0 \hfill & 0 \hfill & 0 \hfill \\
\end{array} }} \right].
\end{equation}
Then in the rotating frame one finds
\begin{equation}
\label{eq24}
F_{rot,ext}^{\mu \nu } =\Lambda ^\mu _A \Lambda ^\nu _B F_{lab,ext}^{AB} 
=\left[ {{\begin{array}{*{20}c}
 0 \hfill & 0 \hfill & 0 \hfill & 0 \hfill \\
 0 \hfill & 0 \hfill & {\frac{B_0 }{r}} \hfill & 0 \hfill \\
 0 \hfill & {-\frac{B_0 }{r}} \hfill & 0 \hfill & 0 \hfill \\
 0 \hfill & 0 \hfill & 0 \hfill & 0 \hfill \\
\end{array} }} \right]=H_{rot,ext}^{\mu \upsilon } 
\end{equation}
where the last part on the RHS follows from \textbf{E}=\textbf{D} and 
\textbf{B}=\textbf{H} external to the cylinder.

The source Maxwell field equations in 4D notation and Gaussian units with 
zero source terms are
\begin{equation}
\label{eq25}
H_{\,\,\,\,\,\,\,;\mu }^{\mu \nu } =\frac{4\pi }{c}J^\nu =0.
\end{equation}
Then applying the Gauss divergence theorem in 4D one gets the boundary 
condition
\begin{equation}
\label{eq26}
n_\mu H_{ext}^{\mu \upsilon } =n_\mu H_{int}^{\mu \upsilon } 
\end{equation}
where $n_{\mu }$ is a 4-vector in the direction of interest. By taking 
various values for $n_{\mu }$ [such as $n_{\mu }$ = (0,1,0,0)] at the cylinder 
boundaries one readily finds that
\begin{equation}
\label{eq27}
H_{rot,int}^{\mu \upsilon } =H_{rot,ext}^{\mu \upsilon } =\left[ 
{{\begin{array}{*{20}c}
 0 \hfill & 0 \hfill & 0 \hfill & 0 \hfill \\
 0 \hfill & 0 \hfill & {\frac{B_0 }{r}} \hfill & 0 \hfill \\
 0 \hfill & {-\frac{B_0 }{r}} \hfill & 0 \hfill & 0 \hfill \\
 0 \hfill & 0 \hfill & 0 \hfill & 0 \hfill \\
\end{array} }} \right],
\end{equation}
where (\ref{eq27}) is the 4D \textbf{D}$,$\textbf{H} tensor internal to the cylinder in 
rotating coordinates. To find the associated 4D \textbf{E}$,$\textbf{B} tensor, 
use the covariant constitutive relations (\ref{eq10}) and (\ref{eq11}) with $u^{\mu }$ and 
$u_{\mu }$.

For a fixed location in the rotating frame, the four-velocity is
\begin{eqnarray}
\label{eq28}
u^\nu =\frac{dx^\nu }{d\tau }=\frac{1}{\sqrt {1-v^2/c^2} }\frac{dx^\nu 
}{dt}  \nonumber \\*
\quad \quad  =\frac{1}{\sqrt {1-v^2/c^2} }\left[ {{\begin{array}{*{20}c}
 {\textstyle{{d(ct)} \over {dt}}} \hfill \\
 0 \hfill \\
 0 \hfill \\
 0 \hfill \\
\end{array} }} \right]=\gamma \left[ {{\begin{array}{*{20}c}
 c \hfill \\
 0 \hfill \\
 0 \hfill \\
 0 \hfill \\
\end{array} }} \right]
\end{eqnarray}
where $\gamma $ has the usual meaning and $v={\omega}r$. The covariant four vector obtained 
from lowering the index of the contravariant 4-velocity $u^{\nu }$ is
\begin{equation}
\label{eq29}
u_\nu =g_{\nu \alpha } u^\alpha =\left[ {{\begin{array}{*{20}c}
 {-c\sqrt {1-v^2/c^2} } \hfill \\
 0 \hfill \\
 {\frac{r^2\omega }{\sqrt {1-v^2/c^2} }} \hfill \\
 0 \hfill \\
\end{array} }} \right].
\end{equation}
Using (\ref{eq28}) with (\ref{eq11}) having the index $\sigma $ = 3, one finds\footnote{ 
Note that the components of $\varepsilon _{\sigma \lambda \mu \nu }$ for a 
non-orthonormal basis are not all 0 or 1. (See ref. \cite{Misner:1973}, 
Exercise 8.3, p. 207.) However, all such components differ from one by a 
factor equal to the determinant of the transformation from a local 
orthonormal basis to the non-orthonormal basis, and that factor cancels on 
both sides of (\ref{eq11}), making calculations herein using (\ref{eq11}) simpler.}, to 
first order,
\begin{equation}
\label{eq30}
F_{rot,int}^{12} =-F_{rot,int}^{21} \cong \frac{\mu B_0 }{r},
\end{equation}
where the initial equality results from the anti-symmetry of the tensor, and 
from here on we freely interchange numeric and letter indices, i.e., $\mu $ 
= $t,r,\phi $, or $z$ is equivalent to $\mu $ = 0,1,2 or 3, respectively.

Using (\ref{eq29}) with (\ref{eq10}) having the index $\mu $ = 1, one obtains 
\begin{equation}
\label{eq31}
2(H_{rot,int}^{10} u_0 +H_{rot,int}^{12} u_2 )=2\varepsilon \left( 
{F_{rot,int}^{10} u_0 +F_{rot,int}^{12} u_2 } \right)
\end{equation}
or
\begin{eqnarray}
\label{eq32}
 & \frac{B_0 }{r}\frac{r^2\omega }{\sqrt {1-v^2/c^2} } &  \nonumber \\
 &  =\varepsilon \left( 
{F_{rot,int}^{12} \frac{r^2\omega }{\sqrt {1-v^2/c^2} }-F_{rot,int}^{10} 
c\sqrt {1-v^2/c^2} } \right). &
\end{eqnarray}
Using (\ref{eq30}) and dropping higher order terms, one finds
\begin{equation}
\label{eq33}
F_{rot,int}^{10} =-F_{rot,int}^{01} \cong -\left( {\frac{1}{\varepsilon 
}-\mu } \right)\frac{vB_0 }{c}.
\end{equation}
Evaluation of other components in similar fashion shows them to be all zero 
and results in
\begin{equation}
\label{eq34}
F_{rot,int}^{\mu \nu } \cong \left[ {{\begin{array}{*{20}c}
 0 \hfill & {\frac{vB_0 }{c}\left( {\frac{1}{\varepsilon }-\mu } \right)} 
\hfill & 0 \hfill & 0 \hfill \\
 {-\frac{vB_0 }{c}\left( {\frac{1}{\varepsilon }-\mu } \right)} \hfill & 0 
\hfill & {\frac{\mu B_0 }{r}} \hfill & 0 \hfill \\
 0 \hfill & {-\frac{\mu B_0 }{r}} \hfill & 0 \hfill & 0 \hfill \\
 0 \hfill & 0 \hfill & 0 \hfill & 0 \hfill \\
\end{array} }} \right].
\end{equation}
As an aside, if we wished to know what \textbf{E} and \textbf{B }field 
values would be measured within the rotating cylinder, we would lower the 
index in (\ref{eq34}) and calculate physical component values\footnote{ We would 
find, to first order, an axial magnetic field of physical magnitude $\mu 
B_{0}$ and a radial electric field of physical magnitude 
\textit{vB}$_{0}$/$c\varepsilon $.}.

Transforming back to the lab, we get, again to first order
\begin{eqnarray}
\label{eq35}
 & F_{lab,int}^{AB} =\Lambda ^A_\mu \Lambda ^B_\nu F_{rot,int}^{\mu \nu } 
 & \nonumber \\*
 & \cong 
\left[ {{\begin{array}{*{20}c}
 0 \hfill & {\frac{vB_0 }{c}\left( {\frac{1}{\varepsilon }-\mu } \right)} 
\hfill & 0 \hfill & 0 \hfill \\
 {-\frac{vB_0 }{c}\left( {\frac{1}{\varepsilon }-\mu } \right)} \hfill & 0 
\hfill & {\frac{\mu B_0 }{R}} \hfill & 0 \hfill \\
 0 \hfill & {-\frac{\mu B_0 }{R}} \hfill & 0 \hfill & 0 \hfill \\
 0 \hfill & 0 \hfill & 0 \hfill & 0 \hfill \\
\end{array} }} \right]. &
\end{eqnarray}
Lowering the index to get the physically relevant covariant form\footnote{ 
We do this step here (and at the beginning of the section) merely to remain 
consistent with the steps laid out in the general methodology of sections 
\ref{subsec:general} and \ref{subsec:overview}. As noted, 
covariant physical components in orthogonal coordinate systems, such as the 
lab, equal contravariant physical components.} yields
\begin{eqnarray}
\label{eq36}
 & F_{\scriptsize \begin{array}{l}
 AB \\ 
 lab,int \\ 
 \end{array}} =G_{A\Gamma } G_{B\Omega } F_{lab,int}^{\Gamma \Omega }
&  \nonumber \\*
  & \cong 
\left[ {{\begin{array}{*{20}c}
 0 \hfill & {-\frac{vB_0 }{c}\left( {\frac{1}{\varepsilon }-\mu } \right)} 
\hfill & 0 \hfill & 0 \hfill \\
 {\frac{vB_0 }{c}\left( {\frac{1}{\varepsilon }-\mu } \right)} \hfill & 0 
\hfill & {\mu RB_0 } \hfill & 0 \hfill \\
 0 \hfill & {-\mu RB_0 } \hfill & 0 \hfill & 0 \hfill \\
 0 \hfill & 0 \hfill & 0 \hfill & 0 \hfill \\
\end{array} }} \right] &
\end{eqnarray}
It is interesting to note that the physical component for the $A=1, B=2$ (i.e., Z 
direction) component above is $\mu B_{0}$, as might be expected. More 
importantly, the electric field in the radial direction in coordinate 
components is the $A=1, B=0$ component above. Hence, the experimentally measured value 
for the radial electric field in the lab is the physical component value
\begin{eqnarray}
\label{eq37}
E_{\scriptsize \begin{array}{l}
 R \\ 
 lab,int \\ 
 measured \\ 
 \end{array}} =F_{\scriptsize \begin{array}{l}
 \hat {1}\hat {0} \\ 
 lab,int \\ 
 \end{array}} =\sqrt {-G^{00}} \sqrt {G^{11}} F_{\scriptsize \begin{array}{l}
 10 \\ 
 lab,int \\ 
 \end{array}}  \nonumber  \\*
\cong \frac{vB_0 }{c}\left( {\frac{1}{\varepsilon }-\mu } 
\right).
\end{eqnarray}
This is the Wilson and Wilson result.

It is important to note how this result hinges on the covariant constitutive 
equations (\ref{eq10}) and (\ref{eq11}) as expressed in an NTO frame. In (\ref{eq31}) to (\ref{eq33}) we get 
the unexpected result that $F^{10}\ne 0$, even though $H^{10}=0$. This is 
due to the NTO nature of the rotating frame. That is, the lowering operation 
performed by $g_{\alpha \beta }$ (with off diagonal terms) yields a $u_{\nu }$ 
having a non-zero $\nu $=2 (i.e. $\phi $ direction) component [see (\ref{eq29})], 
even though $u^{\nu }=\gamma (c$,0,0,0)$^{T}$. This non-zero term in 
$u_{\nu }$ manifests in (\ref{eq10}) [i.e., in (\ref{eq31})] in just the right way to give us 
two extra terms in (\ref{eq32}) and result in (\ref{eq33}).

\section{THE ROENTGEN AND EICHENWALD RESULT}
\label{sec:mylabel2}
With an external electric field in the lab of $E_{0}$ in the axial (i.e., 
$Z)$ direction, and the same transformation steps and boundary condition 
equations used above for deriving the Wilson and Wilson result, one finds, 
internal to the cylinder in the rotating frame,
\begin{equation}
\label{eq38}
H_{rot,int}^{\mu \upsilon } =H_{rot,ext}^{\mu \upsilon } =\left[ 
{{\begin{array}{*{20}c}
 0 \hfill & 0 \hfill & 0 \hfill & {E_0 } \hfill \\
 0 \hfill & 0 \hfill & 0 \hfill & 0 \hfill \\
 0 \hfill & 0 \hfill & 0 \hfill & {-\textstyle{\omega \over c}E_0 } \hfill 
\\
 {-E_0 } \hfill & 0 \hfill & {\textstyle{\omega \over c}E_0 } \hfill & 0 
\hfill \\
\end{array} }} \right].
\end{equation}
Using $\sigma $ = 1 in covariant constitutive relation (\ref{eq11}) yields
\begin{equation}
\label{eq39}
F_{rot,int}^{23} =\mu H_{rot,int}^{23} =-\frac{\omega }{c}E_0 .
\end{equation}
From (\ref{eq10}) with $\mu $=3, $u_{\mu }$ of (\ref{eq29}), and (\ref{eq39}) one then finds
\begin{equation}
\label{eq40}
F_{rot,int}^{30} \cong \mu H_{rot,int}^{30} =-\frac{E_0 }{\varepsilon },
\end{equation}
where we have again dropped higher order terms. Other values for indices 
$\sigma $ and $\mu $ in (\ref{eq10}) and (\ref{eq11}) result in all other tensor components 
of zero, and hence
\begin{equation}
\label{eq41}
F_{rot,int}^{\mu \upsilon } \cong \left[ {{\begin{array}{*{20}c}
 0 \hfill & 0 \hfill & 0 \hfill & {\frac{E_0 }{\varepsilon }} \hfill \\
 0 \hfill & 0 \hfill & 0 \hfill & 0 \hfill \\
 0 \hfill & 0 \hfill & 0 \hfill & {-\textstyle{\omega \over c}\mu E_0 } 
\hfill \\
 {-\frac{E_0 }{\varepsilon }} \hfill & 0 \hfill & {\textstyle{\omega \over 
c}\mu E_0 } \hfill & 0 \hfill \\
\end{array} }} \right].
\end{equation}
Lowering the indices, one obtains the covariant form, to first order
\begin{eqnarray}
\label{eq42}
  & F_{\scriptsize \begin{array}{l} \mu \upsilon \\ 
 rot,int \\ 
 \end{array}} =g_{\mu \alpha } g_{\upsilon \beta } F_{rot,int}^{\alpha \beta 
}  &  \nonumber \\*
   &  \cong \left[ {{\begin{array}{*{20}c}
 0 \hfill & 0 \hfill & 0 \hfill & {-\frac{E_0 }{\varepsilon }} \hfill \\
 0 \hfill & 0 \hfill & 0 \hfill & 0 \hfill \\
 0 \hfill & 0 \hfill & 0 \hfill & {\textstyle{v \over c}r\left( 
{\frac{1}{\varepsilon }-\mu } \right)E_0 } \hfill \\
 {\frac{E_0 }{\varepsilon }} \hfill & 0 \hfill & {-\textstyle{v \over 
c}r\left( {\frac{1}{\varepsilon }-\mu } \right)E_0 } \hfill & 0 \hfill \\
\end{array} }} \right]. & 
\end{eqnarray}
The physical component for the radial direction in the rotating frame is
\begin{eqnarray}
\label{eq43}
B_{\hat {r}} =F_{\scriptsize \begin{array}{l}
 \hat {2}\hat {3} \\ 
 rot,int \\ 
 \end{array}} =\sqrt {g^{22}} \sqrt {g^{33}} F_{\scriptsize \begin{array}{l}
 23 \\ 
 rot,int \\ 
 \end{array}}  \nonumber \\*
\cong \frac{1}{r}F_{\scriptsize \begin{array}{l}
 23 \\ 
 rot,int \\ 
 \end{array}} \cong \left( {\frac{1}{\varepsilon }-\mu } 
\right)\frac{v}{c}E_0 ,
\end{eqnarray}
which for permeability $\mu $ = 1 is the RE result. (See ref. 
\cite{Pellegrini:1995}, p. 700, paragraph after eq. (\ref{eq27}).)

\section{SUMMARY AND CONCLUSIONS}
\label{sec:summary}
A tensor analysis method with very general applicability to mechanics and 
electrodynamics in any non-orthonormal basis vector coordinate system has 
been presented. Application of that method focused on one type of 
non-time-orthogonal (NTO) coordinate frame, the rotating frame, and correct 
predictions were obtained for the Wilson and Wilson and Roentgen/Eichenwald 
experiments.

Essential aspects of the method comprise i) use of the physically relevant 
contravariant or covariant form of vectors/tensors, ii) conversion of 
physical (measured) component values to generalized coordinate component 
values prior to tensor analysis, iii) use of generalized covariant 
constitutive equations during tensor analysis, and iv) conversion after 
tensor analysis of coordinate component answers to physical components for 
comparison with experiment.

The author cautions that we have not shown that the fashionable co-moving 
local Lorentz frame analysis method\cite{Ref:3} and the generalized tensor 
method presented herein are equivalent for NTO frames. While they yield 
identical results in many NTO cases, including those treated herein, they do 
not do so in all. See the author's prior work\cite{Klauber:1} for a 
summary of differences and for logic supporting the generalized tensor 
method as the preferable approach.

\begin{appendix}
\section{PHYSICAL CHARACTER OF VECTOR FORMS}
Since
\begin{equation}
\label{eq44}
dx_\mu =g_{\mu \upsilon } dx^\upsilon 
\end{equation}
if $g_{\mu \nu }$ has $g_{01}\ne $0, then the $x^{0}$ (time) axis is not 
orthogonal to the $x^{1}$ spatial axis, and (with $g_{02}=g_{03}$ = 0)
\begin{equation}
\label{eq45}
dx_0 =g_{00\,} dx^0+g_{01\,} dx^1.
\end{equation}
Hence, unlike \textit{dx}$^{0}$, the component \textit{dx}$_{0}$ represents not a displacement 
purely through time, but an amalgam of displacement through both time 
(\textit{dx}$^{0})$ and space (\textit{dx}$^{1})$.
If $g_{01}$ were to equal zero (time orthogonal 
to space), then \textit{dx}$_{0}$ would comprise a displacement through time only.  

In terms of basis vectors, the \textbf{e}$_{0}$ basis vector is aligned with 
the time axis, and in the time-orthogonal case shares the same line of 
action as its associated basis one-form\cite{Ref:4} 
\bm{$\omega $} $^{0}$. Hence $dx^0{\rm {\bf e}}_0 =dx_0$\bm{$\omega $} $^{0}$ and 
both $dx ^{0}$ and \textit{dx}$_{0}$ represent only time (and no space) displacements.

In the more general case where the \textbf{e}$_{0}$ (time) basis vector is 
not orthogonal to the \textbf{e}$_{i}$ (space) basis vectors, then 
\bm{$\omega $}$^{0}$ does not share the same line of action as 
\textbf{e}$_{0}$ and, $dx_0 $\bm{$\omega $} $^0 \,=dx^0{\rm {\bf e}}_0 +dx^i{\rm {\bf 
e}}_i \,$. Hence, in this (NTO) case \textit{dx}$_{0}$ represents displacement in both 
space and time (i.e., along \bm{$\omega $}$^{0}$, not \textbf{e}$_{0})$, 
whereas \textit{dx}$^{0}$ represents displacement only in time.
Thus, \textit{dx}$_{0}$ and \textit{dx}$^{0}$ can not be considered equivalent in any conceptual 
or physical sense. They do not represent the same physical entity.

\section{PHYSICAL COMPONENTS}
\label{sec:appendix}
Consider an arbitrary vector \textbf{v} in a 2D space
\begin{equation}
\label{eq46}
{\rm {\bf v}}=v^1{\rm {\bf e}}_1 +v^2{\rm {\bf e}}_2 =v^{\hat {1}}{\rm {\bf 
\hat {e}}}_1 +v^{\hat {2}}{\rm {\bf \hat {e}}}_2 ,
\end{equation}
where \textbf{e}$_{i}$ are coordinate basis vectors and ${\rm {\bf \hat 
{e}}}_i$ are unit length (non-coordinate) basis vectors pointing in the same 
respective directions. That is,
\begin{equation}
\label{eq47}
{\rm {\bf \hat {e}}}_i =\frac{{\rm {\bf e}}_i }{\vert {\rm {\bf e}}_i \vert 
}=\frac{{\rm {\bf e}}_i }{\sqrt {{\rm {\bf e}}_{\underline{i}} \cdot {\rm 
{\bf e}}_{\underline{i}} } }=\frac{{\rm {\bf e}}_i }{\sqrt 
{g_{\underline{i}\underline{i}} } },
\end{equation}
where underlining implies no summation. Note that \textbf{e}$_{1}$ and 
\textbf{e}$_{2}$ here do not, in general, have to be orthogonal. Note also, 
that physical components are those associated with unit length basis vectors 
and hence are represented by indices with carets in (\ref{eq46}).

Substituting (\ref{eq47}) into (\ref{eq46}), one readily obtains
\begin{equation}
\label{eq48}
v^{\hat {i}}=\sqrt {g_{\underline{i}\underline{i}} } v^i.
\end{equation}
Relation (\ref{eq48}) between physical and coordinate components is valid locally in 
curved, as well as flat, spaces and can be extrapolated as summarized in 
Section \ref{subsec:physical} to 4D general relativistic 
applications, to higher order tensors, and to covariant 
components\cite{Ref:1988}.

\section{SURFACE CURRENT MODEL NTO ANALYSIS}
\label{sec:mylabel3}
In PS\cite{Pellegrini:1995} Section II, p. 697, magnetization in the 
rotating frame is modeled by an equivalent surface 3D current density [see 
PS eq (6)]
\begin{equation}
\label{eq49}
{\rm {\bf J}}_S ={\rm {\bf M}}\,\times {\rm {\bf \hat {n}}}
\end{equation}
or
\begin{equation}
\label{eq50}
J_{S\;\hat {i}} =\varepsilon _{ijk} M^jn^k=\left[ {{\begin{array}{*{20}c}
 0 \hfill \\
 {\frac{\mu -1}{4\pi }B_0 } \hfill \\
 0 \hfill \\
\end{array} }} \right]=\left[ {{\begin{array}{*{20}c}
 0 \hfill \\
 {\sigma _{rot} } \hfill \\
 0 \hfill \\
\end{array} }} \right],
\end{equation}
where $M^{j}$ is magnetization in the axial direction, $n^{k}$ is a unit 
vector pointing inward (in the radial direction) from the cylinder outer 
surface, $J_{S\hat {\phi }} =\sigma _{rot} $ is the current density on the 
inner surface in the circumferential (i.e., the $\phi )$ direction, and 
$-J_{S\hat {\phi }} =-\sigma _{rot} $ is the current density on the outer 
surface.

Note that (\ref{eq50}) is a ``3-vector'' in terms of physical components. We must 
convert it to coordinate components in 4-vector form, and then raise the 
index to obtain the contravariant form. We then transform this four-vector 
back to the lab frame. The physical zeroth component of this lab four-vector 
is the effective surface charge that would be experimentally measured in the 
lab.

From (\ref{eq50}) and the definition of physical components
\begin{eqnarray}
\label{eq51}
\frac{(\mu -1)}{4\pi }B_0 \;\mbox{=}\;\;\mbox{phys comp }J_{S\;\hat {\phi }} 
=\sqrt {g^{\phi \phi }} J_{S\;\phi } \nonumber \\*
=\frac{\sqrt {1\;-\;v^2/c^2} 
}{r}J_{S\;\phi } 
\end{eqnarray}
where, as before, $v=\omega r.$ Solving the LHS and RHS of (\ref{eq51}) for the 
coordinate component $J_{S\phi }$ allows us to write the covariant 
four-vector as
\begin{equation}
\label{eq52}
J_{S\,\mu } =\left[ {{\begin{array}{*{20}c}
 0 \hfill \\
 0 \hfill \\
 {\frac{(\mu -1)}{4\pi }\frac{rB_0 }{\sqrt {1-v^2/c^2} }} \hfill \\
 0 \hfill \\
\end{array} }} \right].
\end{equation}
The contravariant surface current four-vector is then
\begin{eqnarray}
\label{eq53}
 & J_S ^\mu =g^{\mu \nu }J_{S\;\nu } &  \nonumber \\*
 & = \left[ {{\begin{array}{*{20}c}
 {-1} \hfill & 0 \hfill & {\textstyle{\omega \over c}} \hfill & 0 \hfill \\
 0 \hfill & 1 \hfill & 0 \hfill & 0 \hfill \\
 {\textstyle{\omega \over c}} \hfill & 0 \hfill & 
{\textstyle{{(1\;\;-\;v^2/c^2)} \over {r^2}}} \hfill & 0 \hfill \\
 0 \hfill & 0 \hfill & 0 \hfill & 1 \hfill \\
\end{array} }} \right]\left[ {{\begin{array}{*{20}c}
 0 \hfill \\
 0 \hfill \\
 {\frac{(\mu -1)}{4\pi }\frac{rB_0 }{\sqrt {1-v^2/c^2} }} \hfill \\
 0 \hfill \\
\end{array} }} \right] & \nonumber \\*
 &= \left[ {{\begin{array}{*{20}c}
 {\frac{(\mu -1)}{4\pi }\frac{v}{c}\frac{B_0 }{\sqrt {1-v^2/c^2} }} \hfill 
\\
 0 \hfill \\
 {\sqrt {1-v^2/c^2} \frac{(\mu -1)}{4\pi }\frac{B_0 }{r}} \hfill \\
 0 \hfill \\
\end{array} }} \right]. 
\end{eqnarray}
Note how the off diagonal term in the NTO metric $g^{\mu \nu }$ results in a 
$\mu $=0 term that is not zero, and effectively ``creates'' charge. 
Transforming (\ref{eq53}) to the lab, we get
\begin{eqnarray}
\label{eq54}
J_{\scriptsize \begin{array}{l}
 S \\ 
 lab \\ 
 \end{array}}^{\;\,A} =\Lambda ^A_\beta \,J_S^{\;\,\beta } =\left[ 
{{\begin{array}{*{20}c}
 1 \hfill & 0 \hfill & 0 \hfill & 0 \hfill \\
 0 \hfill & 1 \hfill & 0 \hfill & 0 \hfill \\
 {\textstyle{\omega \over c}} \hfill & 0 \hfill & 1 \hfill & 0 \hfill \\
 0 \hfill & 0 \hfill & 0 \hfill & 1 \hfill \\
\end{array} }} \right]\;\left[ {{\begin{array}{*{20}c}
 {\frac{(\mu -1)}{4\pi }\frac{v}{c}\frac{B_0 }{\sqrt {1-v^2/c^2} }} \hfill 
\\
 0 \hfill \\
 {\sqrt {1-v^2/c^2} \frac{(\mu -1)}{4\pi }\frac{B_0 }{r}} \hfill \\
 0 \hfill \\
\end{array} }} \right]  \nonumber \\*
 \cong \quad \left[ {{\begin{array}{*{20}c}
 {\frac{(\mu -1)}{4\pi }\frac{v}{c}B_0 } \hfill \\
 0 \hfill \\
 {\frac{(\mu -1)}{4\pi }\frac{B_0 }{R}} \hfill \\
 0 \hfill & \\
\end{array} }} \right]
\end{eqnarray}
where higher order terms were dropped on the RHS, and where we note, as did 
PS, that the transformation (\ref{eq12}) to the lab does not create charge.

Physical charge density in the lab to first order is then
\begin{equation}
\label{eq55}
\sigma _{lab} =J_{\scriptsize \begin{array}{l}
 S \\ 
 lab \\ 
 \end{array}}^{\;\,\hat {0}} =\sqrt {-G_{00} } J_{\scriptsize \begin{array}{l}
 S \\ 
 lab \\ 
 \end{array}}^{\;\,0} =\frac{(\mu -1)}{4\pi }\frac{v}{c}B_0 ,
\end{equation}
which equals (7) in PS and, again, yields the Wilson and Wilson result.

\end{appendix}


\begin{thebibliography}{6}
\bibitem{Crater:1994} H. W. Crater, Am. J. Phys. \textbf{62} (\ref{eq10}), 923 (1994)
\bibitem{Ridgely:1999} C. T. Ridgely, Am. J. Phys. \textbf{67}(\ref{eq5}), 414 (1999).
\bibitem{Klauber:1998} R. D. Klauber, Found. Phys. Lett. \textbf{11}(\ref{eq5}), 405 (1998), Sect. 4.3.3 and 4.3.4, pp. 426-429.
\bibitem{Jackson:1975} J. D. Jackson, \textit{Classical Electrodynamics} (John Wiley, New York, 1975), Sect. 11.9, pp. 550-551.
\bibitem{Misner:1973} C. W. Misner, K. S. Thorne, and J. A. Wheeler, \textit{Gravitation} (Freeman, New York, 1973), Chap. 3.
\bibitem{Ref:1988} Ref. \cite{Misner:1973}, see p. 37 and, for example, eq. (31.5) on p. 821; D. Savickas, Am. J. Phys., \textbf{70}(\ref{eq8}), 798; I.S. Sokolnikoff, \textit{Tensor Analysis} (Wiley {\&} Sons, 1951), pp. 8, 122-127, 205; G.E. Hay, \textit{Vector and Tensor Analysis}, (Dover, 1953), pp 184-186; A. J. McConnell, \textit{Application of Tensor Analysis} (Dover, 1947), pp. 303-311; C. E. Pearson, \textit{Handbook of Applied Mathematics} (Van Nostrand Reinhold, 1983 2$^{nd}$ ed.), pp. 214-216; M. R. Spiegel, \textit{Schaum's Outline of Vector Analysis} (Schaum), p. 172; R. C. Wrede, \textit{Introduction to Vector and Tensor Analysis,} (Dover, 1972), pp. 234-235; Malvern, L.E., \textit{Introduction to the Mechanics of a Continuous Medium} (Prentice-Hall, Englewood Cliffs, NJ, 1969), App. I, Sec. 5, pp. 606-613; Y.C. Fung, \textit{Foundations of Solid Mechanics} (Prentice-Hall, Englewood Cliffs, NJ, 1965), pp. 52-53, 111-115; A. C. Eringen, \textit{Nonlinear Theory of Continuous Media (}McGraw-Hill, NY, 1962), pp. 437-439; T. J. Chung, \textit{Continuum Mechanics} (Prentice Hall, Englewood Cliffs, NJ, 1988), pp. 40-53, 246-251.
\bibitem{Ref:1971} Ref. \cite{Misner:1973}, pp. 204-207, 210, and 239, and A.C. Eringen, \textit{Continuum Physics: Volume 1 Mathematics,} (Academic Press, NY and London, 1971), pp. 65-66.
\bibitem{Ref:1} Ref. \cite{Ref:1988}, Fung, p. 53.
\bibitem{Landau:1984} L. Landau and E. Lifshitz, \textit{Electrodynamics of Continuous Media} (Pergamon, New York, 1984), 2$^{nd}$ ed, pp. 260-263.
\bibitem{Klauber:1} R. D. Klauber, in \textit{Relativity in Rotating Frames}, edited by A. van der Merwe (Kluwer Academic Publishers, The Netherlands, in press).
\bibitem{Klauber:1999} R. D. Klauber, Am. J. Phys. \textbf{67}(\ref{eq2}), 158 (1999).
\bibitem{Klauber:2003} R. D. Klauber, Found. Phys. Lett. \textbf{16}(\ref{eq5}) (2003) (to be published).
\bibitem{Ref:2} Ref. \cite{Klauber:1998}, Section 4.1.
\bibitem{Wilson:1913} M. Wilson and H.A. Wilson, \textit{Proc. R.. Soc. London}, Ser. A: \textbf{89}, 99 (1913).
\bibitem{Pellegrini:1995} G. N. Pellegrini and A. R. Swift, Am. J. Phys. \textbf{63}(\ref{eq8}), 694 (1995).
\bibitem{Hertzberg:2001} J. B. Hertzberg, S. R. Bickman, M. T. Hummon, D. Krause, Jr., S. K. Peck, and L. R. Hunter, Am. J. Phys. \textbf{69} (\ref{eq6}), 648 (2001).
\bibitem{Weber:1997} T. A. Weber, Am. J. Phys. \textbf{65}(\ref{eq10}), 946 (1997).
\bibitem{Burrows:1997} M. L. Burrows, Am. J. Phys. \textbf{65} (\ref{eq9}), 929 (1997).
\bibitem{Ridgely:1998} C. T. Ridgely, Am. J. Phys. \textbf{66} (\ref{eq2}), 114 (1998).
\bibitem{Roentgen:1} W. C. Roentgen, Ann. Phys. (Leipzig) \textbf{35,} 264 (1888); \textbf{40}, 93 (1890).
\bibitem{Eichenwald:1904} A. Eichenwald, Ann. Phys. \textbf{11}, 421 (1903); \textbf{13}, 919 (1904).
\bibitem{Ref:3} Ref. \cite{Weber:1997} among many others.
\bibitem{Ref:4} Ref. \cite{Misner:1973}, section 2.7, pp. 60-62.
\end{thebibliography}
\end{document}